\RequirePackage{filecontents}

\RequirePackage{snapshot}
\documentclass[aps,pra,reprint,showpacs,secnumarabic,floatfix,superscriptaddress,longbibliography]{preprint}

\renewcommand\documentclass[2][]{}
\let\origparagraph=\paragraph
\AtBeginDocument{%
  \let\paragraph=\origparagraph
}
\makeatother
\documentclass[reprint,aps,pra,showpacs,floatfix,superscriptaddress,balancelastpage,longbibliography]{revtex4-1}

\usepackage[T1]{fontenc}
\usepackage{mathtools}
\usepackage{textcomp}
\usepackage{txfonts}
\usepackage{tgtermes}
\usepackage[altg]{qtxmath}
\usepackage{MnSymbol}
\makeatletter
\newcommand{\raisemath}[1]{\mathpalette{\raisem@th{#1}}}
\newcommand{\raisem@th}[3]{\raisebox{#1}{$#2#3$}}
\def\hbar{\ensuremath{\mathrlap{\raisemath{-0.1ex}{{\mathchar'26}}}\mkern-1mu h}}
\makeatother

\usepackage{microtype}
\usepackage{graphicx}
\usepackage{braket}
\graphicspath{{figures/}}
\usepackage[breaklinks=true,
pdfauthor={Q. Z. Lv, Heiko Bauke, Q. Su, C. H. Keitel, and R. Grobe},
pdftitle={Bosonic pair creation and the Schiff-Snyder-Weinberg effect}]{hyperref}

\newcommand*{\op}[1]{{\hat#1}}
\newcommand*{\D}{\mathrm{d}}
\newcommand*{\e}{\mathrm{e}}
\newcommand*{\I}{\mathrm{i}}

\newcommand*{\htrans}[1]{#1^{\mathsf{\dagger}}}
\renewcommand*{\vec}[1]{\boldsymbol{#1}}
\newcommand*{\mat}[1]{\boldsymbol{#1}}
\newcommand*{\conj}[1]{#1^*}
\newcommand*{\commut}[2]{\mathchoice{\left[#1,#2\right]}{[#1,#2]}{[#1,#2]}{[#1,#2]}}

\DeclareMathOperator{\IM}{Im}

\begin{document}

\title{Bosonic pair creation and the Schiff-Snyder-Weinberg effect}

\author{Q. Z. Lv}
\affiliation{Max-Planck-Institut f{\"u}r Kernphysik, Saupfercheckweg~1,
  69117~Heidelberg, Germany} 
\affiliation{Beijing National Laboratory for Condensed Matter Physics,
  Institute of Physics, Chinese Academy of Sciences, Beijing 100190,
  China}
\affiliation{Intense Laser Physics Theory Unit and Department of
  Physics, Illinois State University, Normal, Illinois 61790-4560, USA}
\author{Heiko Bauke}
\email{heiko.bauke@mpi-hd.mpg.de}
\affiliation{Max-Planck-Institut f{\"u}r Kernphysik, Saupfercheckweg~1,
  69117~Heidelberg, Germany}
\author{Q. Su}
\affiliation{Intense Laser Physics Theory Unit and Department of
  Physics, Illinois State University, Normal, Illinois 61790-4560, USA}
\author{C. H. Keitel}
\affiliation{Max-Planck-Institut f{\"u}r Kernphysik, Saupfercheckweg~1, 
  69117~Heidelberg, Germany}
\author{R. Grobe}
\affiliation{Intense Laser Physics Theory Unit and Department of
  Physics, Illinois State University, Normal, Illinois 61790-4560, USA}

\begin{abstract}
  Interactions between different bound states in bosonic systems can
  lead to pair creation.  We study this process in detail by solving the
  Klein-Gordon equation on space-time grids in the framework of
  time-dependent quantum field theory.  By choosing specific external
  field configurations, two bound states can become pseudodegenerate,
  which is commonly referred to as the Schiff-Snyder-Weinberg
  effect. These pseudodegenerate bound states, which have complex energy
  eigenvalues, are related to the pseudo-Hermiticity of the Klein-Gordon
  Hamiltonian. In this work, the influence of the Schiff-Snyder-Weinberg
  effect on pair production is studied.  A generalized
  Schiff-Snyder-Weinberg effect, where several pairs of pseudodegenerate
  states appear, is found in combined electric and magnetic fields.  The
  generalized Schiff-Snyder-Weinberg effect likewise triggers pair
  creation.  The particle number in these situations obeys an
  exponential growth law in time enhancing the creation of bosons, which
  cannot be found in fermionic systems.
\end{abstract}


\date{January 22, 2016}

\maketitle

\allowdisplaybreaks

\section{Introduction}
\label{sec:intro}

The possibility to create particle-antiparticle pairs from the vacuum in
a strong external force field has sparked considerable interest among
the theoretical
\cite{Schutzhold:Gies:etal:2008:Dynamically_Assisted,Ruf:Mocken:etal:2009:Pair_Production,Bulanov:Mur:etal:2010:Multiple_Colliding,Cheng:Su:etal:2009:Creation_of,Di-Piazza:Muller:etal:2012:Extremely_high-intensity,Kohlfurst:Gies:etal:2014:Effective_Mass,Pike:Mackenroth:etal:2014:photon-photon_collider,Wollert:Bauke:etal:2015:Spin_polarized}
as well as the experimental communities
\cite{Burke:Field:etal:1997:Positron_Production,Ahmad:Austin:etal:1997:Search_for,Bamber:Boege:etal:1999:Studies_of,Kasper:etal:2015:Schwinger_pair_production}.
The number of created particles during the interaction is associated
directly with transitions from (initially occupied) states of negative
energy to those of the positive energy.  There are two mechanisms that
can cause these upward transitions.  The first one requires the external
force to be time dependent.  Here the force field could be provided by
one or several electromagnetic radiation pulses.  In fact, several
experimental efforts are presently planned worldwide to develop lasers
with extremely high intensities that could breakdown the vacuum
\cite{McNeil:Thompson:2010:X-ray_free-electron,Mourou:Fisch:etal:2012:Exawatt-Zettawatt_pulse}.

The second mechanism of pair creation is based on static fields that can
temporarily lead to effective spectral degeneracies among the energy
states during the interaction.  The most frequently studied case
involves the degeneracy of two continua, as is characteristic for the
Schwinger mechanism, predicted by Sauter
\cite{Sauter:1931:Ueber_Verhalten}, Heisenberg and Euler
\cite{Heisenberg:Euler:1936:Folgerungen_aus}, and Schwinger
\cite{Schwinger:1951:On_Gauge}.  This scenario is also directly related
to the so-called Klein paradox
\cite{Klein:1929:Reflexion_von,Klein:1927:Elektrodynamik_und,Hansen:Ravndal:1981:Kleins_Paradox,Dombey:1999:Seventy_years,Holstein:1999:Strong_field,Krekora:Su:etal:2004:Klein_Paradox},
where a sufficiently steep and large potential barrier can trigger the
pair creation.  In the early 1970s it was predicted
\cite{Zeldovich:Popov:1972:Electronic_structure,Muller:1983:Quantum_Electrodynamics,Muller:Peitz:etal:1972:Solution_of,Muller:Rafelski:etal:1972:Electron_shells}
that the degeneracy between a potential's ground state and the lower
continuum can also trigger pair creation.

It seems therefore natural to expect that the creation of particle pairs
should also become possible due to a degeneracy of two discrete states.
However, the generic behavior of two coupled discrete states is
generally characterized by avoided crossings
\cite{Landau:Lifshitz:1981:Quantum_Mechanics}.  A recent work
\cite{Liu:Lv:etal:2015:Pair_creation} examined the Dirac equation for a
field configuration characterized by a combination of an attractive well
and a repulsive well, which could support simultaneously bound states
for particles as well as antiparticles.  By decreasing the spatial
separation between the two neighboring wells it was possible to couple
the electronic and positronic ground states with each other and to
examine the effect of the corresponding avoided crossing (quasi
degeneracy) on the change in the total particle number.  Due to the
unavoidable occurrence of avoided crossings it is not possible to
construct a field configuration for a fermionic system for which an
exact degeneracy can be found in the discrete spectrum of the Dirac
equation.

For the Klein-Gordon equation there are field configurations, for which
a pseudodegeneracy of discrete states can occur, in contrast to the
Dirac equation.  This was first pointed out by Snyder and Weinberg
\cite{Snyder:Weinberg:1940:Stationary_States} and Schiff, Snyder, and
Weinberg \cite{Schiff:Snyder:etal:1940:On_Existence} in 1940, who
discovered (rather counterintuitively) that a sufficiently strong and
finite-range potential can support simultaneous bound states for the
particle as well as the antiparticle.  Later the restriction of the
compact spatial support was generalized to potentials that only have to
be short range
\cite{Villalba:Rojas:2006:Bound_states,Rafelski:Fulcher:etal:1978:Fermions_and,Popov:1971:Position_Production}.
The Schiff-Snyder-Weinberg effect was related to pair creation by
various authors
\cite{Arbuzov:Rochev:1972:On_production,Popov:1971:Position_Production,Fulling:1976:Varieties_of,Villalba:Rojas:2006:Bound_states}.
Here we are going to study bosonic pair creation under the
Schiff-Snyder-Weinberg effect systematically in the framework of quantum
field theory.

The paper is organized as follows.  In Sec.~\ref{sec:Klein_Gordon} we
discuss general properties of the Klein-Gordon equation in its
Feshbach-Villars representation, in particular its spectrum and the
properties of the eigenvectors, which leads us to the concept of
pseudodegeneracy and the Schiff-Snyder-Weinberg effect, which is studied
in Sec.~\ref{sec:SSW_effect}.  Section~\ref{sec:pair_creation} examines
the creation of bosonic particle-antiparticle pairs induced by the
pseudodegeneracy of the Klein-Gordon equation and shows that the
pair-creation rate can be obtained from the imaginary parts of the
Hamilton operator's eigenvalues.  Pseudodegeneracies between discrete
states do not necessarily require external binding fields. In fact, a
suitable superposition of a non-binding electric field and a magnetic
field can also lead to Schiff-Snyder-Weinberg-like states that
eventually become pseudodegenerate, as shown in
Sec.~\ref{sec:pair_creation_magnetic}.  How a possible back reaction of
the created particles on the electromagnetic environment modifies the
pair-creation dynamics is investigated in Sec.~\ref{sec:backreaction}.
Finally, we complete this paper with an outlook and conclusions in
Sec.~\ref{sec:Conclusions}.

\section{The Klein-Gordon equation}
\label{sec:Klein_Gordon}

Before considering pair creation within a quantum field theoretical
framework let us firstly take a look at the spectrum of the Klein-Gordon
equation for a single particle.  A bosonic quantum particle of mass $m$
and charge $q$ interacting with the electromagnetic potentials $\vec
A(\vec r, t)$ and $\phi(\vec r, t)$ can be described by the Klein-Gordon
equation in the Feshbach-Villars representation
\cite{Feshbach:Villars:1958:Elementary_Relativistic,Greiner:1997:Relativistic_Quantum_Mechanics,Wachter:2011:Relativistic_Quantum_Mechanics}:
\begin{multline}
  \label{eq:Klein-Gordon-Schroedinger}
  \I\hbar\frac{\partial \Psi(\vec r, t)}{\partial t}
   = \op H_\mathrm{KG}\Psi(\vec r, t) = \\
  \left(
    \frac{\sigma_3+\I\sigma_2}{2m}
    \left( -\I\hbar\vec\nabla - q \vec A(\vec r, t) \right)^2
    + q \phi(\vec r, t) + \sigma_3 mc^2
  \right)\Psi(\vec r, t) \,,
\end{multline}
where the Pauli matrices $\sigma_1$, $\sigma_2$, and $\sigma_3$ are
defined as
\begin{equation}
  \label{eq:Pauli_matrices}
  \sigma_1 = \begin{pmatrix} 0&1\\ 1&0 \end{pmatrix}\,{,}\quad
  \sigma_2 = \begin{pmatrix} 0&-\I\\ \I&0 \end{pmatrix}\,{,}\quad
  \sigma_3 = \begin{pmatrix} 1&0\\ 0&-1 \end{pmatrix} \,.
\end{equation}

In contrast to the Schr{\"o}dinger equation, the Hamilton operator $\op
H_\mathrm{KG}$, which determines the time evolution of the two-component
wave function $\Psi(\vec r, t)$, is not Hermitian and the corresponding
time-evolution operator is not unitary.  The Klein-Gordon Hamilton
operator is, however, $\sigma_3$ pseudo-Hermitian and its time-evolution
operator is $\sigma_3$ pseudounitary
\cite{Mostafazadeh:2004:Pseudounitary_operators,Mostafazadeh:2010:Pseudo-Hermitian_representation}.
A linear operator $\op H$ is called \mbox{$\op\eta$ pseudo Hermitian} if
there is a Hermitian operator $\op\eta$ such that $\op H$ equals its
so-called $\op\eta$ pseudoadjoint operator $\op H^\#=\op\eta^{-1}\op
H^\dagger\op\eta$, i.\,e.,
\begin{equation}
  \op\eta^{-1}\op H^\dagger\op\eta =  \op H\,.
\end{equation}

Hermitian Hamilton operators have real eigenvalues, their eigenfunctions
form an orthonormal basis, and the vectors in the dual space which are
associated to the eigenfunctions are given by the complex conjugated
eigenfunctions.  Pseudo-Hermitian operators such as the Klein-Gordon
Hamilton operator, however, may have complex eigenvalues and the
eigenfunctions and the dual functions are not just the complex
conjugates of each other.

For simplicity let us assume that the Hilbert space has a finite
dimension $d$.  Then the pseudo-Hermitian Hamilton operator can be
represented by a finite matrix $\mat H$ with some set of right
eigenvectors $\vec\psi_i$ (the so-called kets, represented by column
vectors) and the corresponding left eigenvectors $\vec\varphi_i$ (the
so-called bras or dual vectors, represented by row vectors).  The left
and right eigenvectors form a biorthogonal system, i.\,e.,
$\vec\varphi_i\vec\psi_j\ne 0$ if and only if $i=j$.  Note that because
left and right eigenvectors are defined only up to a multiplicative
complex constant, one can always find left and right eigenvectors such
that $\vec\varphi_i\vec\psi_i=1$. Corresponding left and right
eigenvectors $\vec\varphi_i$ and $\vec\psi_i$ have the same eigenvalue
$\mathcal{E}_i$. If the number of left and right eigenvectors equals $d$
then each of both sets of vectors spans the whole Hilbert space and any
column vector $\vec\Psi$ can be expandend into the basis of right
eigenvectors by multiplying with the left eigenvectors, i.\,e.,
\begin{equation}
  \label{eq:expansion}
  \vec\Psi = \sum_{i=1}^d 
  \frac{\vec\varphi_i\vec\Psi}{\vec\varphi_i\vec\psi_i}\vec\psi_i\,.
\end{equation}
Thus, the biorthogonal system $\vec\psi_i$ and $\vec\varphi_i$ plays the
same role as the orthogonal systems do in the Hermitian case.  The
notion of biorthogonal vectors and the results, which follow in the
reminder of this section, can be generalized to biorthogonal function
systems in case of infinite dimensional Hilbert spaces
\cite{Pell:1911:Biorthogonal_systems,Bari:1951:Biorthogonal,Gohberg:Krein:1969:Introduction,Mostafazadeh:2002:Pseudo-Hermiticity_versus,Mostafazadeh:2002:Pseudo-Hermiticity_versus_II,Mostafazadeh:2002:Pseudo-Hermiticity_versus_III,Curtright:Mezincescu:2007:Biorthogonal_quantum,Brody:2014:Biorthogonal_quantum}.

For each left eigenvector $\vec\varphi_i$,
\begin{equation}
  \htrans{(\vec\varphi_i\mat H)} = 
  \htrans{(\mathcal{E}_i\vec\varphi_i)} = 
  \conj{\mathcal{E}_i}\htrans{\vec\varphi_i} 
\end{equation}
holds, but due to pseudo-Hermiticity also
\begin{equation}
  \htrans{(\vec\varphi_i\mat H)} = 
  \htrans{\mat H}\htrans{\vec\varphi_i} = 
  \mat\eta\mat H{\mat\eta}^{-1}\htrans{\vec\varphi_i} \,,
\end{equation}
where $\mat\eta$ is in the case of the Klein-Gordon equation a
block-diagonal matrix with $\sigma_3$ replicated on the diagonal and
$\mat\eta={\mat\eta}^{-1}=\htrans{\mat\eta}$.  Consequently
\begin{equation}
  \label{eq:left_right}
  \mat H{\mat\eta}^{-1}\htrans{\vec\varphi_i} =
  \conj{\mathcal{E}_i}{\mat\eta}^{-1}\htrans{\vec\varphi_i}  \,,
\end{equation}
and therefore ${\mat\eta}^{-1}\htrans{\vec\varphi_i}$ is a right
eigenvector of $\mat H$ with the eigenvalue $\conj{\mathcal{E}_i}$.
Thus, eigenvalues of $\mat H$ must be real or come in pairs of
$\mathcal{E}_i$ and its complex conjugate $\conj{\mathcal{E}_i}$,
i.\,e., there is some $\mathcal{E}_j$ such that
$\mathcal{E}_j=\conj{\mathcal{E}_i}$.  We call eigenstates with
$\mathcal{E}_j=\conj{\mathcal{E}_i}$ pseudodegenerate because the
corresponding energy eigenvalues have equal real parts but different
imaginary parts.  Furthermore, it follows from Eq.~\eqref{eq:left_right}
for a pair of pseudodegenerate states that the right eigenvector of
state $j$ is related to the left eigenvector of state $i$ by
$\vec\psi_j={\mat\eta}^{-1}\htrans{\vec\varphi_i}$ or equivalently
$\vec\varphi_i=\htrans{\vec\psi_j}\mat\eta$.  If $\mathcal{E}_i$ is
real, however, it follows from Eq.~\eqref{eq:left_right} that
${\mat\eta}^{-1}\htrans{\vec\varphi_i}$ is a right eigenvector
corresponding to the left eigenvector ${\vec\varphi_i}$, i.\,e.,
$\vec\psi_i={\mat\eta}^{-1}\htrans{\vec\varphi_i}$ or equivalently
$\vec\varphi_i=\htrans{\vec\psi_i}\mat\eta$.

Therefore, one may introduce the pseudo inner product $(\vec\Psi_1,
\vec\Psi_2)_\eta$ between two quantum states $\vec\Psi_1$ and
$\vec\Psi_2$ as
\begin{equation}
  \label{eq:pseudo_inner}
  \left(\vec\Psi_1, \vec\Psi_2\right)_\eta =
  \htrans{\vec\Psi_1}\mat{\eta}\vec\Psi_2\,,
\end{equation}
which is commonly applied in the context of the Klein-Gordon equation
\cite{Feshbach:Villars:1958:Elementary_Relativistic,Greiner:1997:Relativistic_Quantum_Mechanics,Wachter:2011:Relativistic_Quantum_Mechanics},
and the expansion \eqref{eq:expansion} may be written as
\begin{equation}
  \label{eq:expansion_bar}
  \vec\Psi = \sum_{i=1}^d 
  \frac{\htrans{\vec\psi_i}\mat\eta\vec\Psi}{\htrans{\vec\psi_i}\mat\eta\vec\psi_i}\vec\psi_i\,,
\end{equation}
provided that the Hamilton matrix $\mat H$ has a real spectrum, e.\,g.,
in the free-particle case.

It is crucial, however, to realize that the pseudo inner product
\eqref{eq:pseudo_inner} is problematic, in particular if the spectrum
contains complex eigenvalues.  The first problem is that states cannot
always be normalized to plus one with respect to the inner product
\eqref{eq:pseudo_inner}.  Secondly, the denominator in
Eq.~\eqref{eq:expansion_bar} becomes
$\htrans{\vec\psi_i}\mat\eta\vec\psi_i=0$ if $\mathcal{E}_i$ is complex,
which follows from $\htrans{\vec\psi_i}\mat\eta=\vec\varphi_j$ and the
orthogonality of $\vec\varphi_j$ and $\vec\psi_i$.  Therefore, the
scalar product should be formed only between left and right eigenvectors
as in Eq.~\eqref{eq:expansion}.

The time evolution of an eigenstate $\vec\psi_i$ of $\mat H$ is
\begin{equation}
  \label{eq:psi_i_t}
  \vec\psi_i(t)=\e^{-\I\mathcal{E}_it/\hbar}\vec\psi_i \,.
\end{equation}
Thus, if $\mathcal{E}_i$ is complex then the components of
$\vec\psi_i(t)$ shrink or grow exponentially.  However, the scalar
product between $\vec\psi_i(t)$ and its dual vector $\vec\varphi_i(t)$ is
constant, i.\,e.,
\begin{equation}
  \vec\varphi_i(t)\vec\psi_i(t) = \vec\varphi_i\vec\psi_i = \mathrm{const.}
\end{equation}
This result follows from Eq.~\eqref{eq:psi_i_t} and
\begin{equation}
  \vec\varphi_i(t) =
  \htrans{\vec\psi_j(t)}\mat\eta =
  \htrans{(\e^{-\I\mathcal{E}_jt/\hbar}\vec\psi_j)}\mat\eta =
  \e^{\I\mathcal{E}_it/\hbar}\htrans{\vec\psi_j}\mat\eta =
  \e^{\I\mathcal{E}_it/\hbar}\vec\varphi_i \,,
\end{equation}
where $\vec\psi_i$ and $\vec\psi_j$ denote a pair of pseudodegenerate
states.  Obviously,
$\vec\varphi_i(t)\vec\psi_i(t)=\vec\varphi_i\vec\psi_i$ holds for
eigenstates with real energy eigenvalue, too.  Furthermore, if the
Hamilton matrix $\mat H$ has a real spectrum then each general column
vector $\vec\Psi(t)$, which evolves under $\mat H$, satisfies
\begin{equation}
  \htrans{\vec\Psi(t)}\mat\eta\vec\Psi(t) =
  \htrans{\vec\Psi(0)}\mat\eta\vec\Psi(0) = \mathrm{const.}
\end{equation}

\section{The Schiff-Snyder-Weinberg effect}
\label{sec:SSW_effect}

\begin{figure}
  \centering
  \includegraphics[scale=0.475]{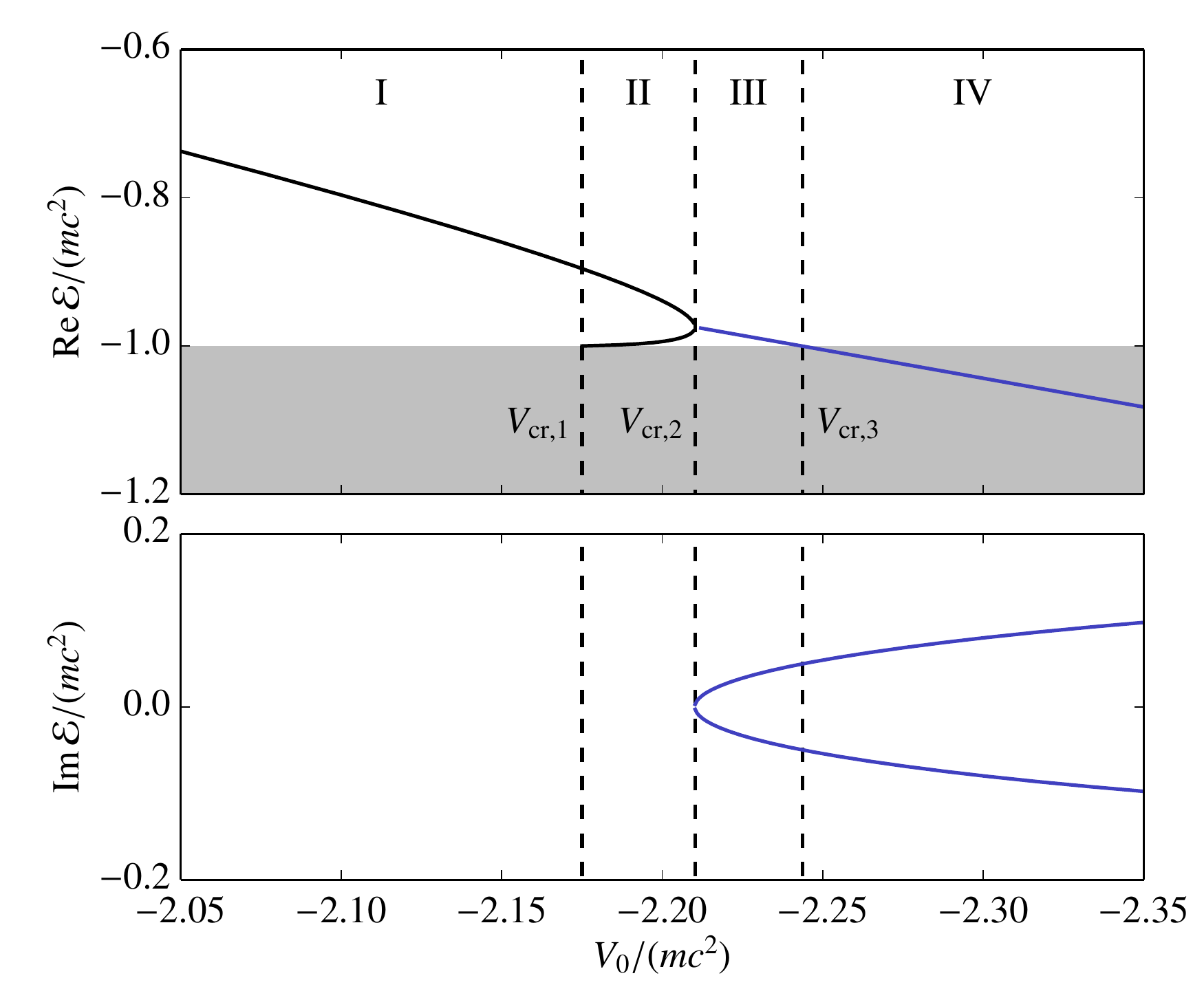}
  \caption{(Color online) The Schiff-Snyder-Weinberg effect for a
    one-dimensional smooth box potential
    $q\phi(x)=V_0/2\times(\mbox{$\tanh((x+l/2)/w)$}-\mbox{$\tanh((x-l/2)/w)$})$
    with the box width $l=2.2\,\lambda_\mathrm{C}$ and
    $w=0.2\,\lambda_\mathrm{C}$, where $\lambda_\mathrm{C}$ denotes the
    particle's Compton length.  The solid lines show the real part and
    the imaginary part (if nonzero) of some bound-state energy
    eigenvalues as a function of the potential strength $V_0$, the gray
    shaded area represents the negative-energy continuum.  See main text
    for further details.}
  \label{fig:SSW}
\end{figure}

Back in 1940 Schiff, Snyder, and Weinberg investigated in their
pioneering work \cite{Schiff:Snyder:etal:1940:On_Existence} the unusual
properties of the Klein-Gordon equation due to its non-Hermiticity.
They determined the bound states of the equation for a deep square well
potential for varying potential strengths.  For a narrow but not too
deep potential the Klein-Gordon equation features only bound states with
energy close to but below the positive-continuum threshold at
$mc^2$. The deeper and therefore more attractive the potential is, the
smaller the eigenenergies of these positive-continuum bound states are;
see also Fig.~\ref{fig:SSW}.  At a first critical potential strength
$V_\mathrm{cr,1}$ a further bound state emerges above but close to the
negative-continuum threshold at $-mc^2$.  We call this state the
negative-continuum bound state.  Contrary to other bound states its
energy eigenvalue grows as the potential gets deeper.  As the new state
appears at the border to the negative continuum one may relate it to an
antiparticle.  This, however, leads to the paradox situation that the
strong potential, which is repulsive for an antiparticle with charge
$-q$, features a bound antiparticle state.  Thus, if the potential depth
grows beyond $V_\mathrm{cr,1}$ the energy values move away from the
continuum thresholds and approach each other and finally two energy
eigenvalues merge at a second critical potential strength
$V_\mathrm{cr,2}$.  Thus, the states become degenerate at the point
$V_\mathrm{cr,2}$, which is not possible in one-dimensional quantum
systems with finite potentials within the nonrelativistic Schr{\"o}dinger
theory \cite{kar:Parwani:2007:Can_degenerate}.  Note that the degenerate
states at $V_\mathrm{cr,2}$ are linearly independent in contrast to
one-dimensional complex $PT$-symmetric potentials, where linear
independence is lost, when eigenvalue curves intersect
\cite{Ahmed:Ghosh:etal:2015:Accidental_crossings}.  Beyond the critical
point $V_\mathrm{cr,2}$ the energy eigenvalues are complex and the
states become pseudodegenerate.  The real part of the energy eigenvalues
decreases further with growing potential depth and enters finally the
negative-energy continuum, thus there is a discrete bound state embedded
in a continuum of states
\cite{Stillinger:Herrick:1975:Bound_states,Lv:Liu:etal:2013:Noncompeting_Channel}.\enlargethispage{2ex}

In summary, one can distinguish four parameter regimes.  In regime~I the
bound states move down towards the negative continuum.  In regime~II an
antiparticle bound state is present. The two bound states become pseudo
degenerate in regimes~III and~IV.  They have complex energy eigenvalues
with the real part lying in the gap between the positive-energy and
negative-energy continua for the regime~III, while the real part is in
the negative-energy continuum for the regime~IV.

This phenomenon of merging of two bound states into a pair of pseudo
degenerate states with complex energy values is called the
Schiff-Snyder-Weinberg effect.  The asymptotic limit of the
Schiff-Snyder-Weinberg effect, that is, for infinite walls, and its
connection to the well-known Klein paradox has been studied by Fulling
\cite{Fulling:1976:Varieties_of}.  In 1970s, Popov came to the
conclusion that the Schiff-Snyder-Weinberg effect is inherent to
short-range interactions and should not be expected for long-range
potentials \cite{Popov:1971:Position_Production}.  Schroer and Swieca
\cite{Schroer:Swieca:1970:Indefinite_Metric} have constructed a formal
quantization of a charged Klein-Gordon field with strong stationary
external interactions including the complex energy modes.  An
application of this quantization to a free scalar field with a tachyonic
mass, i.\,e., $m^2<0$, was given by Schroer
\cite{Schroer:1971:Quantization_of}.  In all these studies, physicists
argued that after the merging of the two bound states, the eigenstates
with complex energy have zero norm, that there is no Fock-like
representation, and postulated a breakdown of the vacuum and a breakdown
of the particle interpretation of quantum field theory.  As we have
shown in the previous section, however, zero norm does not occur if the
scalar product of the correct left and right eigenvectors is utilized.
Thus, there is no apparent reason to postulate a breakdown of the
Klein-Gordon theory in the presence of complex energy values.

\section{Pair creation process under the pseudodegeneracy}
\label{sec:pair_creation}

As discussed in Sec.~\ref{sec:SSW_effect}, pseudodegenerate bound
states appear in the parameter regime~III of strong localized
potentials.  This is a special effect in bosonic systems as it can never
be found in fermionic systems.  This section will discuss bosonic pair
creation via an external scalar potential within a quantum field
theoretical framework.  In particular, we will investigate the role of
pseudodegenerate states and the occurrence of complex energy levels in
this process.

The quantum field operator of a many-particle system can be expressed as
an integral or a sum (in case of a discretized Hamiltonian) over the
simultaneous eigenstates of the free-particle Klein-Gordon Hamiltonian
and the momentum operator:
\begin{equation}
  \label{eq:field_operator}
  \op\Psi(\vec r, t) = 
  \sum_{\vec p} \op b_{\vec p}^+(t)\varphi_{\vec p}^+(\vec r) +
  \sum_{\vec p} \htrans{\op b_{\vec p}^-{}}(t)\varphi_{\vec p}^-(\vec r)\,.
\end{equation}
Here, $\varphi_{\vec p}^+(\vec r)$ denotes a free-particle state with
positive energy and momentum eigenvalue $\vec p$ and correspondingly
$\varphi_{\vec p}^-(\vec r)$ denotes a free-particle state with negative
energy.  The operators $\op b_{\vec p}^+(t)$ and $\op b_{\vec p}^-(t)$
denote the annihilation operators for the particle and antiparticle,
respectively, with momentum $\vec p$.  Together with the
respective creation operators $\htrans{\op b_{\vec p}^+{}}(t)$ and
$\htrans{\op b_{\vec p}^-{}}(t)$ they satisfy the commutator relations
of bosonic annihilation and creation operators
\begin{equation}
  \label{eq:creation_comm}
  \commut{\op b_{\vec p}^\pm(t)}{\htrans{\op b_{\vec p'}^\pm{}}(t)}=
  \delta_{\vec p, \vec p'}
  \,,
\end{equation}
where $\delta_{\vec p, \vec p'}$ denotes a Kronecker delta.  The
particle and antiparticle annihilation operators $\op b_{\vec p}^+(t)$
and $\op b_{\vec p}^-(t)$ both satisfy the Heisenberg equation
\begin{equation}
  \I\hbar \frac{\partial \op b_{\vec p}^\pm(t)}{\partial t} =
  \commut{\op b_{\vec p}^\pm(t)}{\op H} \,,
\end{equation}
where $\op H$ denotes the quantum-field-theoretical Hamiltonian of a
particle coupled to an external field.  This can be expressed in terms
of the first-quantization Klein-Gordon Hamiltonian $\op{H}_\mathrm{KG}$:
\begin{equation}
  \op H = \int 
  \htrans{\op\Psi}(\vec r, t)\sigma_3 \op{H}_\mathrm{KG}\op\Psi(\vec r, t)
  \,\D^3 r \,.
\end{equation}
It should be noted that, while this Hamiltonian fully accounts for the
interaction of the particle due to the external field through the
minimum coupling principle, which is implemented in
$\op{H}_\mathrm{KG}$, it neglects all internal forces between particles
and antiparticles.  

The field operator $\op\Psi(\vec r, t)$ defined in
Eq.~\eqref{eq:field_operator} satisfies the Schr{\"o}dinger-like equation
\begin{equation}
  \I\hbar\frac{\partial\op\Psi(\vec r, t)}{\partial t} =
  \op H_\mathrm{KG} \op\Psi(\vec r, t) \,.
\end{equation}
Consequently, the time-dependent field operator
\eqref{eq:field_operator} may be equivalently expressed as
\begin{equation}
  \label{eq:field_operator_2}
  \op\Psi(\vec r, t) = 
  \sum_{\vec p} \op b_{\vec p}^+\varphi_{\vec p}^+(\vec r, t) +
  \sum_{\vec p} \htrans{\op b_{\vec p}^-{}}\varphi_{\vec p}^-(\vec r, t)\,,
\end{equation}
where $\op b_{\vec p}^+=\op b_{\vec p}^+(0)$ and $\htrans{\op b_{\vec
    p}^-{}}= \htrans{\op b_{\vec p}^-{}}(0)$, and the functions
$\varphi_{\vec p}^+(\vec r, t)$ and $\varphi_{\vec p}^-(\vec r, t)$
denote the solutions of the time-dependent Klein-Gordon equation
\eqref{eq:Klein-Gordon-Schroedinger} with $\varphi_{\vec p}^+(\vec r)$
and $\varphi_{\vec p}^-(\vec r)$, respectively, as initial conditions at
time $t=0$.  By equating Eqs.~\eqref{eq:field_operator}
and~\eqref{eq:field_operator_2} we find
\begin{equation}
  \op b_{\vec p}^+ (t) =
  \sum_{\vec p'} 
  \op b_{\vec p'}^+\braket{\varphi_{\vec p}^+(\vec r)|\varphi_{\vec p'}^+(\vec r, t)} +
  \htrans{\op b_{\vec p'}^-{}}\braket{\varphi_{\vec p}^+(\vec r)|\varphi_{\vec p'}^-(\vec r, t)}
\end{equation}
and
\begin{equation}
  \htrans{\op b_{\vec p}^-{}} (t) =
  \sum_{\vec p'} 
  \htrans{\op b_{\vec p'}^-{}} \braket{\varphi_{\vec p}^-(\vec r)|\varphi_{\vec p'}^-(\vec r, t)} +
  \op b_{\vec p'}^+\braket{\varphi_{\vec p}^-(\vec r)|\varphi_{\vec p'}^+(\vec r, t)} \,,
\end{equation}
where we have employed the pseudo scalar product
\begin{equation}
  \braket{\varphi_{\vec p}^\pm(\vec r)|\Psi(\vec r)} 
  \int \htrans{\varphi_{\vec p}^\pm{}}(\vec r)\sigma_3\Psi(\vec r)\,\D^3 r \,.
\end{equation}
We can calculate the particles' spatial density $\varrho(\vec r, t)$
from the particle portion of the particle-antiparticle field operator,
which is defined as
\begin{equation}
  \label{eq:field_operator_+}
  \op\Psi^+(\vec r, t) = 
  \sum_{\vec p} \op b_{\vec p}^+(t)\varphi_{\vec p}^+(\vec r) \,,
\end{equation}
via the expectation value of the corresponding density operator with
respect to the vacuum state $\ket{|0}$, i.\,e.,
\begin{equation}
  \varrho(\vec r, t) = 
  \braket{0||\htrans{\op\Psi^+{}}(\vec r, t) \op\Psi^+(\vec r, t)||0} \,.
\end{equation}
After some operator algebra and employing Eq.~\eqref{eq:creation_comm} we
find that $\varrho(\vec r, t)$ can be expressed in terms of the
time-dependent solutions of the single-particle Klein-Gordon equation
\eqref{eq:Klein-Gordon-Schroedinger}, viz.
\begin{equation}
  \label{eq:density}
  \varrho(\vec r, t) = 
  \sum_{\vec p'}\Big|
    \sum_{\vec p}
    \braket{\varphi_{\vec p}^+(\vec r) | \varphi_{\vec p'}^-(\vec r, t)}
    \varphi_{\vec p}^+(\vec r)
    \Big|^2
    \,.
\end{equation}
By integrating Eq.~\eqref{eq:density} over the whole space, one obtains
the total number of the created pairs as
\begin{equation}
  \label{eq:particle_number}
  N(t) = \int \varrho(\vec r, t)\,\D^3 r = 
  \sum_{\vec p'}\sum_{\vec p}
  \left|
    \braket{\varphi_{\vec p}^+(\vec r) | \varphi_{\vec p'}^-(\vec r, t)}
  \right|^2\,.
\end{equation}
These expressions permit us to study the details of the pair-creation
process for various parameters (such as the potential height and width)
by investigating the total number of created particles $N(t)$ and the
spatial and the momentum probability distributions of the created
pairs. For an alternative approach based on in and out states see,
e.\,g.,
Refs.~\cite{Fradkin:Gitman:Shvartsman:1991:Quantum_Electrodynamics,Greiner:Muller:Rafelski:1985:Quantum_Electrodynamics}.
While this approach leads to the same result as the asymptotic in- and
out-state-based S-matrix formalism after the external field is turned
off, our approach permits us to follow the dynamics with space-time
resolution \cite{Cheng:Su:etal:2010:Introductory_review}.

\begin{figure}
  \centering
  \includegraphics[scale=0.475]{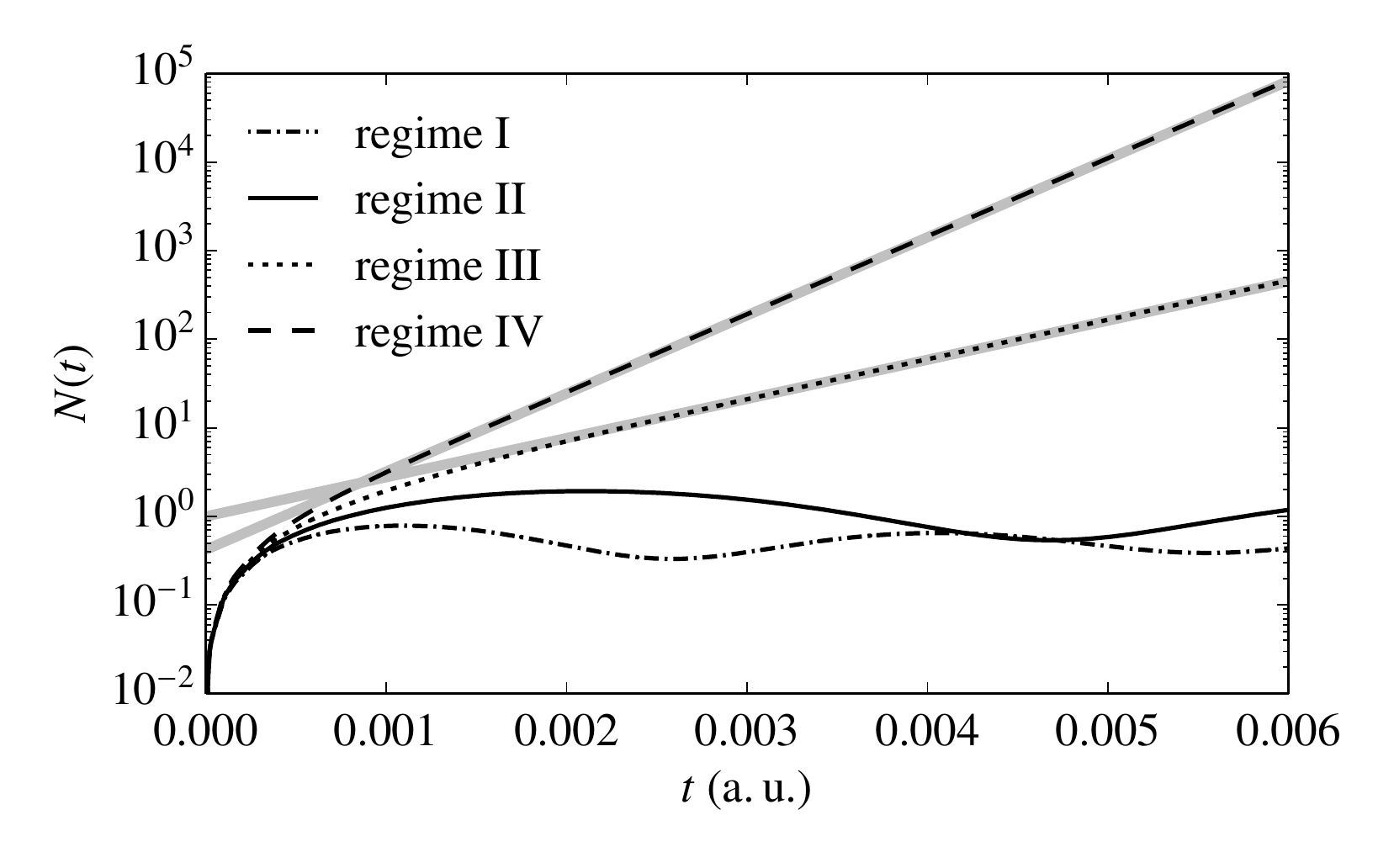}
  \caption{The mean value of the number of created particles $N(t)$ as a
    function of time $t$ for different parameter regimes as indicated in
    Fig.~\ref{fig:SSW} with the same potential and $V_0=-2.17\,mc^2$,
    $V_0=-2.195\,mc^2$, $V_0=-2.22\,mc^2$, and $V_0=-2.25\,mc^2$ for
    regimes I--IV.  In the parameter regimes~I and~II, where all
    eigenenergies are real-valued, the number of created particles
    remains bounded.  In the parameter regimes~III and~IV, however, the
    number of created particles grows exponentially and the growth rate
    is given by twice the absolute value of the imaginary part of the
    pseudodegenerate complex-valued eigenenergies, which appear in
    these regimes.  This exponential growth law is indicated by the
    solid gray lines.  Atomic units (indicated as a.\,u.) are employed.}
  \label{fig:created_particles}
\end{figure}

Figure~\ref{fig:SSW} indicates that the one-dimensional Klein-Gordon
equation may feature complex eigenvalues depending on the parameters and
one may distinguish four parameter regimes with different
characteristics regarding the energy eigenvalues.  The time-dependent
process of pair creation from the initial vacuum and how the
Schiff-Snyder-Weinberg effect affects the pair-creation dynamics may be
studied by utilizing the potential energy
\begin{equation}
  q\phi(x, t)=
  \begin{cases} 
    0 & \text{for $t\le 0$,} \\[1ex]
    \dfrac{V_0}{2}\left( 
      \tanh\dfrac{x+l/2}{w} - \tanh\dfrac{x-l/2}{w}
    \right) & \text{else.}
  \end{cases}
\end{equation}
In Fig.~\ref{fig:created_particles} the mean value of the number of
created particles $N(t)$ is presented as a function of time $t$ for a
parameter set from each of the four parameter regimes.

Parameter regimes~I and~II show qualitatively a completely different
behavior compared to regimes~III and~IV, where pseudodegeneracy is
present.  In the parameter regimes~I and~II the number of created
particles initially grows and then oscillates, i.\,e., it remains
bounded, whereas the number of created particles grows exponentially in
the parameter regimes~III and~IV.  The oscillation of the particle
number in parameter regimes~I and~II is caused by the sudden turn-on of
the external field.  The oscillation frequency times $\hbar$ equals
approximately the energy difference between the lowest
positive-continuum bound state and the negative continuum (regime~I) or
the lowest positive-continuum bound state and the negative-continuum
bound state (regime~II).  Thus, the system behaves like an effective
two-level system.

The lowest positive-continuum bound state and the negative-continuum
bound state are degenerate with respect to the real part of their
energies in the parameter regimes~III and~IV.  In this way, the
positive-energy continuum gets coupled to the negative-energy continuum,
which induces pair creation and an exponential growth of the mean value
of the number of created particles as a function of time.  To calculate
the number of created particles we write Eq.~\eqref{eq:particle_number}
as
\begin{align}
  \nonumber N(t) & = 
  \sum_{\vec p'}\sum_{\vec p} \Big|
    \braket{\varphi_{\vec p}^+(\vec r) | \varphi_{\vec p'}^-(\vec r, t)}
  \Big|^2 \\
  \nonumber & = 
  \sum_{\vec p'}\sum_{\vec p} \Big| 
    \braket{\varphi_{\vec p}^+(\vec r) | 
      \exp(-\I \op H_\mathrm{KG}t/\hbar) | \varphi_{\vec p'}^-(\vec r)}
  \Big|^2 \\
  \nonumber & =
  \sum_{\vec p'}\sum_{\vec p} \Big|
    \Braket{\varphi_{\vec p}^+(\vec r) | 
      \exp(-\I \op H_\mathrm{KG}t/\hbar) 
      \smash{\sum_{\mathcal{E}}}\Ket{\mathcal{E}}\!\Bra{\mathcal{E}}
      \varphi_{\vec p'}^-(\vec r)}
  \Big|^2 \\
  \label{eq:N_t}
  & =
  \sum_{\vec p'}\sum_{\vec p} \Big| 
    \Braket{\varphi_{\vec p}^+(\vec r) | 
      \smash{\sum_{\mathcal{E}}}\exp(-\I \mathcal{E}t/\hbar)
      \Ket{\mathcal{E}}\!\Bra{\mathcal{E}} \varphi_{\vec p'}^-(\vec r)}
  \Big|^2 \,, 
\end{align}
where $\bra{\mathcal{E}}$ and $\ket{\mathcal{E}}$ denote all the left
and right eigenstates of the Klein-Gordon Hamiltonian with energy
$\mathcal{E}$. Supposing that the pseudodegenerate states with energies
$\mathcal{E}$ and $\mathcal{E^*}$ (where without loss of generality
$\IM\mathcal{E}>0$) mainly contribute to the pair creation as the
continua are far away from each other we can approximate
Eq.~\eqref{eq:N_t} by
\begin{align}
  \nonumber
  N(t) & \approx \sum_{\vec p'}\sum_{\vec p} 
  \big|\big\langle
  \varphi_{\vec p}^+(\vec r) \big(\! 
  \exp(-\I\mathcal{E}t/\hbar)\Ket{\mathcal{E}}\!\Bra{\mathcal{E}} \\[-1ex]
  \nonumber
  &\hspace{28mm} +
  \exp(-\I\mathcal{E}^*t/\hbar)\Ket{\mathcal{E}^*}\!\Bra{\mathcal{E}^*} 
  \big)\,\varphi_{\vec p'}^-(\vec r) 
  \big\rangle\big|^2 \\[1ex]
  & \approx \exp(2\IM\mathcal{E}\,t)\sum_{\vec p'}\sum_{\vec p}
  \left|
    \braket{\varphi_{\vec p}^+(\vec
      r)\!\ket{\mathcal{E}}\!\bra{\mathcal{E}} \varphi_{\vec
        p'}^-(\vec r)}
  \right|^2\,.
\end{align}
Thus, the growth rate is given by twice the absolute value of the
imaginary part of the pseudodegenerate eigenenergies, which agrees very
well with our numerical calculations as indicated in
Fig.~\ref{fig:created_particles}.  This self-amplified creation process
in bosonic systems can be understood as the bosons obey the so-called
anti-Pauli blocking principle.  As long as the created particles are
localized in the interaction regime, the sequential creation can be
amplified and the particle number shows exponential increase in time.

\section{Generalized Schiff-Synder-Weinberg effect and its influence on
  pair creation}
\label{sec:pair_creation_magnetic}

The addition of a magnetic field to an electric field can lead to new
phenomena in the Schiff-Synder-Weinberg effect.  In the following, we
will consider pair creation at the step potential
\begin{equation}
  \label{eq:phi_ex}
  q \phi(x, t) = \dfrac{V_0}{2}\left( 
      \tanh\dfrac{x}{w_E} + 1
    \right) \,,
\end{equation}
which corresponds to a strong localized electric field.  The electric
field of the potential \eqref{eq:phi_ex} is not able to support bound
states and, therefore, the Schiff-Synder-Weinberg effect cannot occur.
It may, however, feature bound states if a sufficiently strong localized
magnetic field is superimposed perpendicularly to the localized electric
field \cite{Lv:Su:etal:2013:Pair_creation}.  In the magnetic field the
charged particle can undergo bound cyclotron motion.  Such a magnetic
field may be given by the vector potential
\begin{equation}
  \label{eq:A_ex}
  \vec A(x, t) =
  \begin{pmatrix}
    0 \\
    A_y(x, t) \\
    0
  \end{pmatrix} =
  \begin{pmatrix}
    0 \\
    \dfrac{A_0}{2}\left( 
      \tanh\dfrac{x}{w_B} + 1
    \right) \\
    0
  \end{pmatrix}\,.
\end{equation}
As the scalar potential \eqref{eq:phi_ex} and the vector potential
\eqref{eq:A_ex} depend on the $x$~coordinate only, the system is
quasi-one-dimensional and the canonical momenta in the $y$ and $z$
directions are conserved.  Consequently, the three-dimensional system
can be simplified to a set of non-coupled one-dimensional systems with
different canonical momenta $p_y$ and $p_z$.  In the following, we
choose $p_y=qA_0/2$ and $p_z=0$, which lets the kinematic momentum
components along the $y$ and $z$ direction vanish at $x\approx 0$.
These particular parameters are motivated by the fact that particles
with low kinetic energy are favored in pair creation in strong
electromagnetic fields.

\begin{figure}
  \centering
  \includegraphics[scale=0.475]{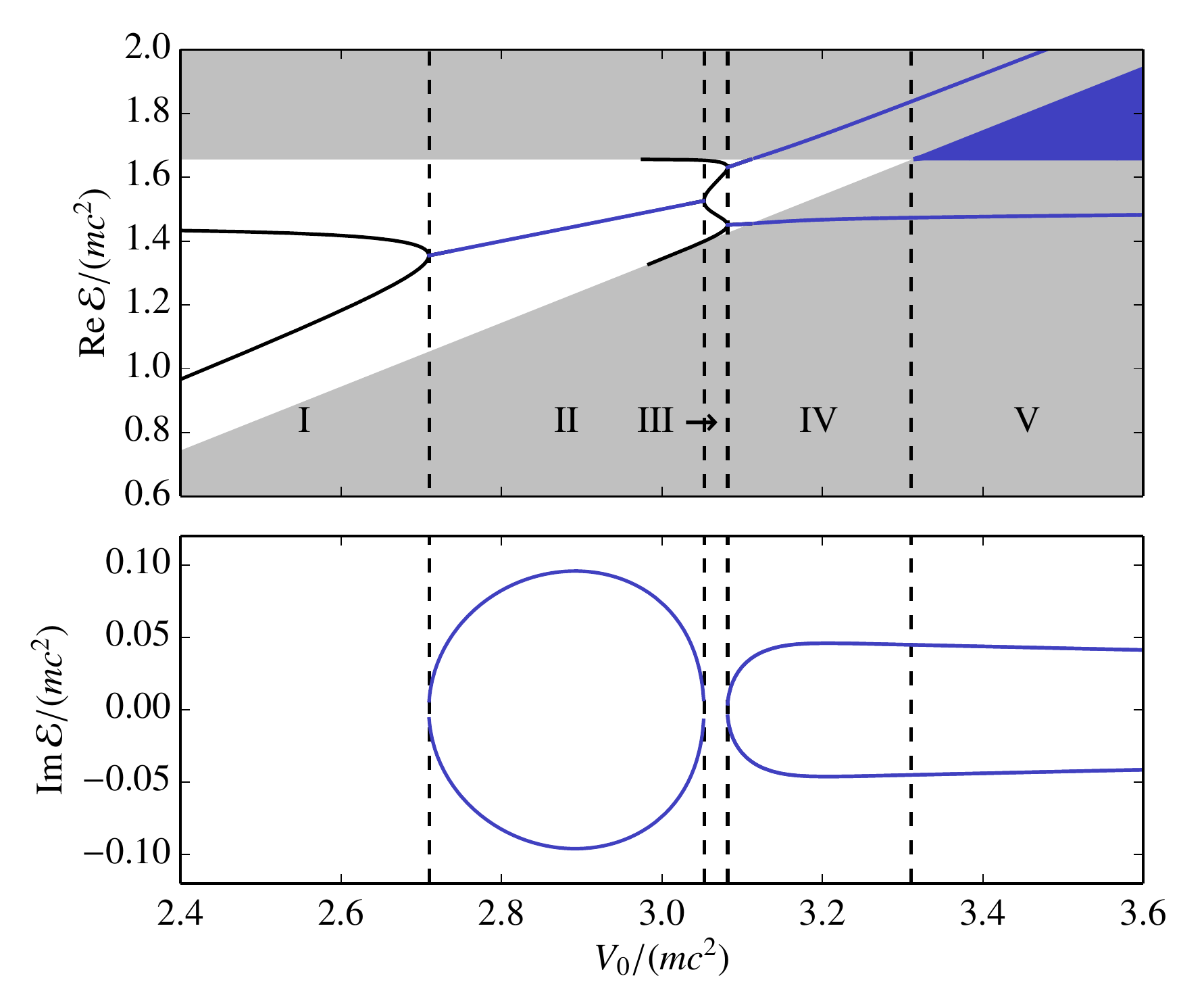}
  \caption{(Color online) Spectrum of the one-dimensional Klein-Gordon
    Hamiltonian \eqref{eq:Klein-Gordon-Schroedinger_1d} with
    $p_y=qA_0/2$ and $p_z=0$, $A_0=1.2\,w_Bmc/(\lambda_\mathrm{C} q)$,
    $w_E=0.3\,\lambda_\mathrm{C}$, and $w_B=2.2\,\lambda_\mathrm{C}$,
    where $\lambda_\mathrm{C}$ denotes the Compton wave length.  The
    solid lines show the real part and the imaginary part (if nonzero)
    of some bound-state energy eigenvalues as a function of the
    potential strengths $V_0$; the shaded areas represent the
    negative-energy and the positive-energy continua.  Discrete states
    and states in the overlapping region of the two continua with
    complex energy eigenvalues are marked in blue.}
  \label{fig:SSW3}
\end{figure}

Figure~\ref{fig:SSW3} shows the spectrum of the Hamiltonian of the
one-dimensional Klein-Gordon equation
\begin{multline}
  \label{eq:Klein-Gordon-Schroedinger_1d}
  \I\hbar\frac{\partial \Psi(x, t)}{\partial t}
  = 
  \Bigg(
    \frac{\sigma_3+\I\sigma_2}{2m}
    \Bigg( 
      -\hbar^2\frac{\partial^2}{\partial x^2} + p_y^2 + p_z^2 \\
      - 2 q p_yA_y(x, t) 
      + q^2A_y(x, t)^2
    \Bigg)
    + q \phi(x, t) + \sigma_3 mc^2
  \Bigg)\Psi(x, t) \,.
\end{multline}
Its spectrum features a lower continuum, an upper continuum of states,
and energetically isolated bound states in between.  The lower limit of
the upper continuum is given by
\begin{equation}
  \mathcal{E} = 
  \sqrt{\smash[b]{c^2(p_y^2+p_z^2) + m^2 c^4}} \,,
\end{equation}
while the upper limit of the lower continuum is given by
\begin{equation}
  \mathcal{E} = 
  -\sqrt{\smash[b]{c^2((p_y-qA_0)^2+p_z^2) + m^2 c^4}} + V_0\,.
\end{equation}
If the scalar potential is strong enough, the two continua intersect,
which happens at
\begin{equation}
  \label{eq:V_0_crit}
  V_0 = 
  \sqrt{\smash[b]{c^2(p_y^2+p_z^2) + m^2 c^4}} +
  \sqrt{\smash[b]{c^2((p_y-qA_0)^2+p_z^2) + m^2 c^4}} \,.
\end{equation}
We find, analog to the electric-field-only Schiff-Synder-Weinberg
effect, for not too strong electric potentials bound states with
real-valued energy, indicated as regime~I in Fig.~\ref{fig:SSW3}.  At
some critical potential strength these states become degenerate and
beyond this critical point the states are pseudodegenerate with complex
conjugated energy values (regime~II).  In contrast to the standard
Schiff-Synder-Weinberg effect the pseudodegenerate states do not dive
into the lower continuum if the strength of the scalar potential is
further increased.  Instead the bound states become degenerate again at
some further critical point and the energies become real and separate
again (regime~III).  The increase of the potential strength also induces
two new bound states between the energy continua and finally triggers a
second occurrence of the Schiff-Synder-Weinberg effect.  This means,
beyond some critical point two pairs of states become simultaneously
pseudodegenerate (regime~IV).  Remarkably, the imaginary parts of all
four states have the same absolute value.  These states finally dive
into the upper and lower continua.  At the potential strength given by
Eq.~\eqref{eq:V_0_crit} the two continua overlap.  In this regime,
indicated as regime~V in Fig.~\ref{fig:SSW3}, a whole continuum of
states with complex energy exists.

The observed coalescing followed by anticoalescing of two levels when
increasing the potential strength does not occur in the standard
Schiff-Synder-Weinberg effect as described in Sec.~\ref{sec:SSW_effect}.
We call this new phenomenon therefore the generalized
Schiff-Synder-Weinberg effect.  Note that the Schiff-Synder-Weinberg
effect has been studied recently for a one-dimensional system including
both a scalar as well as a vector potential, which corresponds to a
vanishing magnetic field in this case
\cite{Hassanabadi:Castro:2015:Bound_States}.  For this system the
generalized Schiff-Synder-Weinberg effect could not be observed.  Thus,
one might argue that the presence of a magnetic field is pivotal for the
generalized Schiff-Synder-Weinberg effect.

It is instructive to analyze the wave functions of the bound states of
the system in more detail.  For this purpose the left and right
eigenstates $\varphi(x)$ and $\psi(x)$ (of the discretized version) of
the Hamilton operator in Eq.~\eqref{eq:Klein-Gordon-Schroedinger_1d} are
calculated, where the states are two-component functions.  More
precisely, $\varphi(x)$ and $\psi(x)$ assign to each space point a
two-component complex row vector and a two-component complex column
vector, respectively.  This allows us to define the density
\begin{equation}
  \label{eq:KG_density}
  \rho(x) = \varphi(x)
  \begin{pmatrix}
    1 & 0 \\ 0 & 1
  \end{pmatrix}
  \psi(x)
\end{equation}
and the eigenstates shall be normalized such that
\begin{equation}
  \label{eq:normalization}
  \int \rho(x)\,\D x = 1 \,.
\end{equation}
\begin{figure}
  \centering
  \includegraphics[scale=0.475]{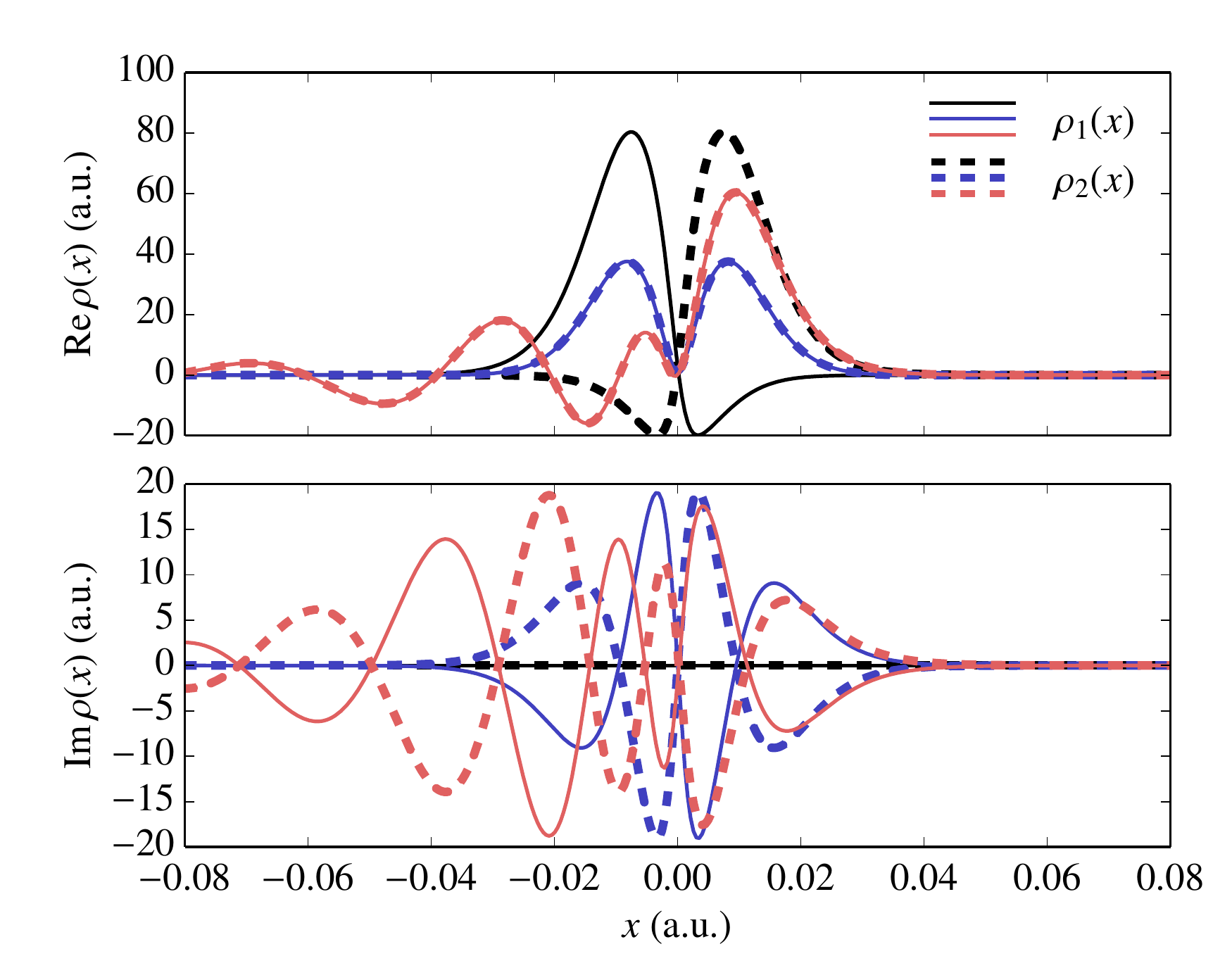}
  \caption{(Color online) Real and imaginary parts of two bound states'
    densities of the Hamilton operator in
    Eq.~\eqref{eq:Klein-Gordon-Schroedinger_1d} denoted by $\rho_1(x)$
    and $\rho_2(x)$ for $V_0=2.6\,mc^2$ (black), $V_0=2.9\,mc^2$ (blue),
    and $V_0=3.2\,mc^2$ (light red).  Other parameters are as for
    Fig.~\ref{fig:SSW3}.  For $V_0=3.2\,mc^2$ there are two pairs of
    pseudodegenerate bound states.  Only the densities of the pair with
    larger real part of the energy eigenvalue are shown.  The densities
    of the other pair of states can be obtained by mirroring the shown
    densities at $x=0$.}
  \label{fig:SSW3b}
\end{figure}%
Note that despite this normalization the density \eqref{eq:KG_density}
may be negative or even complex.  Figure~\ref{fig:SSW3b} shows the real
and imaginary parts of two bound states' densities of the Hamilton
operator in Eq.~\eqref{eq:Klein-Gordon-Schroedinger_1d} for parameters
in regimes~I, II, and~IV of Fig.~\ref{fig:SSW3}.  In regime~I, that is,
where the eigenenergies are real-valued, the densities are real-valued
too.  The bound state that is associated with the upper continuum is
located left from the scalar potential step (solid black line in
Fig.~\ref{fig:SSW3b}), while the bound state that is associated with the
lower continuum is located right from the scalar potential step (dashed
black line in Fig.~\ref{fig:SSW3b}).  This changes if some energy values
become complex as in regime~II.  In this case it is no longer possible
to associate the bound states with one of the two energy continua.  The
densities are complex-valued and the bound states are each located on
both sides of the scalar potential step (solid and dashed blue lines in
Fig.~\ref{fig:SSW3b}).  They occupy the spatial region approximately
from $-w_B$ to $w_B$.  In this way these states bridge states related to
the upper continuum, which are located on the left to the potential step
at $x=0$, and states related to the lower continuum, which are located
on the right to the potential step at $x=0$.  In regime~III the energy
eigenvalues become real again.  In total there are four bound states in
regime~III.  Two of them emerged from the upper and the lower continuum,
respectively, and can be interpreted as particle and antiparticle
states.  The other two states, which lie energetically between them,
emerged from the pseudodegenerate pair of states in the regime~II and
therefore one cannot associate them unambiguously to particle or
antiparticle states.  As a consequence of the generalized
Schiff-Synder-Weinberg effect two pairs of pseudodegenerate states
appear in regime~IV.  The densities of one of these pairs are indicated
in Fig.~\ref{fig:SSW3b} by light red lines.  Similarly to regime~II the
densities are complex-valued but they are no longer symmetric around the
potential step at $x=0$. The densities of the second pair of states (not
shown in Fig.~\ref{fig:SSW3b}) can be obtained by mirroring the
densities of the first pair at $x=0$.

\begin{figure}[t]
  \centering
  \includegraphics[scale=0.475]{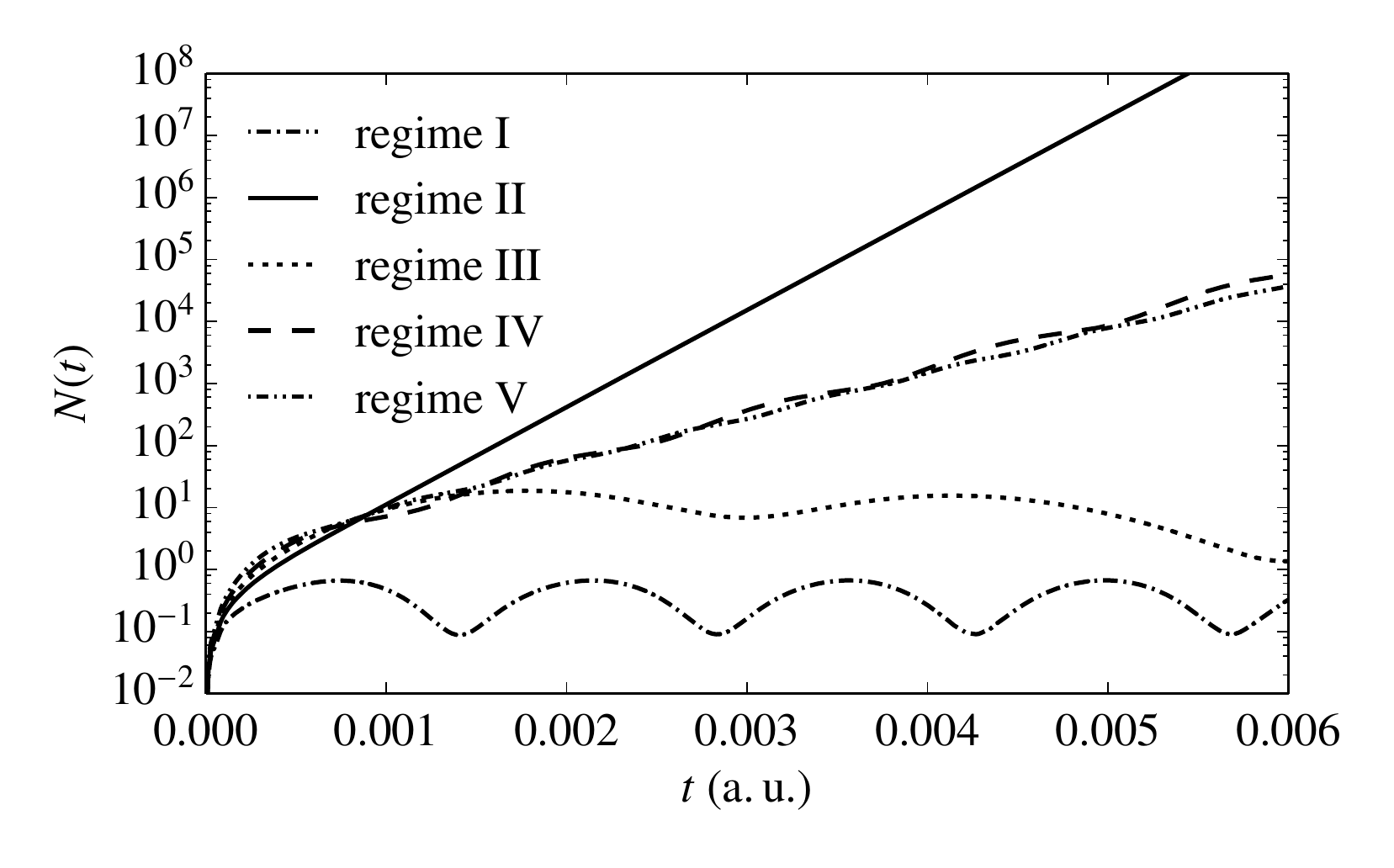}
  \caption{The mean value of the number of created particles $N(t)$ as a
    function of time $t$ for different parameter regimes as indicated in
    Fig.~\ref{fig:SSW3} with the same potential and $V_0=2.6\,mc^2$,
    $V_0=2.\,9mc^2$, $V_0=3.07\,mc^2$, $V_0=3.2\,mc^2$, and
    $V_0=3.4\,mc^2$ for the regimes \mbox{I--V}.  In the parameter
    regimes~I and~III, where all eigenenergies are real valued, the
    number of created particles remains bounded.  In the parameter
    regimes~II, IV, and~V, however, the number of created particles
    grows exponentially and the growth rate is given by twice the
    absolute value of the imaginary part of the pseudodegenerate
    complex valued eigenenergies, which appear in these regimes.}
  \label{fig:created_particles_magnetic}
\end{figure}

The generalized Schiff-Synder-Weinberg effect controlls the
pair-creation process as illustrated in
Fig.~\ref{fig:created_particles_magnetic}.  Depending on the system
parameters and the resulting energy eigenvalues the number of created
articles may grow exponentially or oscillate in time. In regimes~I
and~III all eigenenergies are real-valued and consequently the number of
created particles grows only initially due to the sudden turn-on of the
potentials but then evolves into an oscillatory behavior with a
frequency, which is approximately given by the energy difference of the
two bound states divided by $\hbar$.  This corresponds to regime~II of
the system as described in Sec.~\ref{sec:pair_creation}.  In regime~II
of Fig.~\ref{fig:SSW3}, a single pseudodegenerate pair of bound states
is present and the mean number of created particles grows exponentially
with a rate that is approximately given by twice the absolute value of
the imaginary part of the energy eigenvalues. As a consequence of the
generalized Schiff-Synder-Weinberg effect two pseudodegenerate pairs of
bound states are present in regimes~IV and~V.  Also in these regimes,
the mean number of created particles grows exponentially with a rate
that equals twice the absolute value of the imaginary part of the energy
eigenvalues.  Note that the growth rate can be larger in regime~II than
in regime~IV although the scalar potential is always larger in
regime~IV, as it can be inferred from the imaginary part of energy
eigenvalues shown in Fig.~\ref{fig:SSW3}.  In addition to the
exponential growth of the particle number there is a small oscillation
superimposed to the exponential growth law.  The oscillation frequency
times $\hbar$ equals the difference of the real parts of the energies of
the two pairs of pseudodegenerate bound states.  Also here, the system
behaves like an effective two-level system.  This holds true even in
regime~V, where the two continua of the energy spectrum overlap and the
system has a continuum of pseudodegenerate states.  These scattering
states, however, are not localized around the potential step at $x=0$
and cause a linear growth only of the mean number of created particles,
which is covered by the exponential growth law
\cite{Lv:Liu:etal:2014:Degeneracies_of}.

\section{Back reaction from the created particles}
\label{sec:backreaction}

In the previous sections pair creation was studied in a model where the
electromagnetic field which triggers the pair-creation dynamics is
incorporated via given potentials.  Thus, the electromagnetic field is
not a dynamical variable because a possible back reaction of the created
particles on the external field has been neglected.  Such a back
reaction may eventually stop the seemingly unlimited growth of the
number of created bosons.  In order to account for how the created
particles modify the external field, we introduce a purely
phenomenological model here by considering the energy transfer between
the particles and the field.

The energy of external electric and magnetic fields $\vec{E}(\vec{r})$
and $\vec{B}(\vec{r})$ is with the permittivity of the vacuum
$\varepsilon_{0}$ given by
\begin{equation}
  \label{eq:ini_energy}
  \mathcal{E} = 
  \frac{\varepsilon_{0}}{2} \int \left( 
    \vec{E}(\vec{r})^2 + c^2\vec{B}(\vec{r})^2 
  \right)\,\D^3 r\,.
\end{equation}
The created pairs
have the rest mass energy $2mc^2N(t)$, where $N(t)$ is the number of
pairs.  The created antiparticles are able to escape from the potential;
the created particles, however, remain in the interaction zone.  As
these particles carry charges, they induce an additional electric field
$\vec{E}_\mathrm{in}(\vec{r}, t)$, which reduces the total electric
field and can be obtained by solving the equation
\begin{equation}
  \vec\nabla\vec E_\mathrm{in} (\vec{r},t) = 
  4 \pi \varrho(\vec{r}, t) \,.
\end{equation}
Here $\varrho(\vec{r}, t)$ denotes the spatial distribution of the
created bosons \eqref{eq:density}.  Therefore, the total energy needed
to create the particles is
\begin{equation}
  \label{eq:pair_energy}
  \mathcal{E}_\mathrm{in}(t) = 
  2mc^2N(t) + \frac{\varepsilon_{0}}{2} \int \vec{E}_\mathrm{in}(\vec r, t)^2\,\D^3 r\,.
\end{equation}
The kinetic energy and the magnetic field, which is triggered by the
created bosons, are neglected as the particles are created at rest. Thus,
the energy that is left in the external field after the pair creation is
\begin{equation}
  \label{eq:field_energy}
  \mathcal{E}_\mathrm{ex}(t) =
  \mathcal{E}_{0} - \mathcal{E}_\mathrm{in}(t)\,,
\end{equation}
where $\mathcal{E}_{0}$ denotes the energy of the initial field
configuration.

\begin{figure}
  \centering
  \includegraphics[scale=0.475]{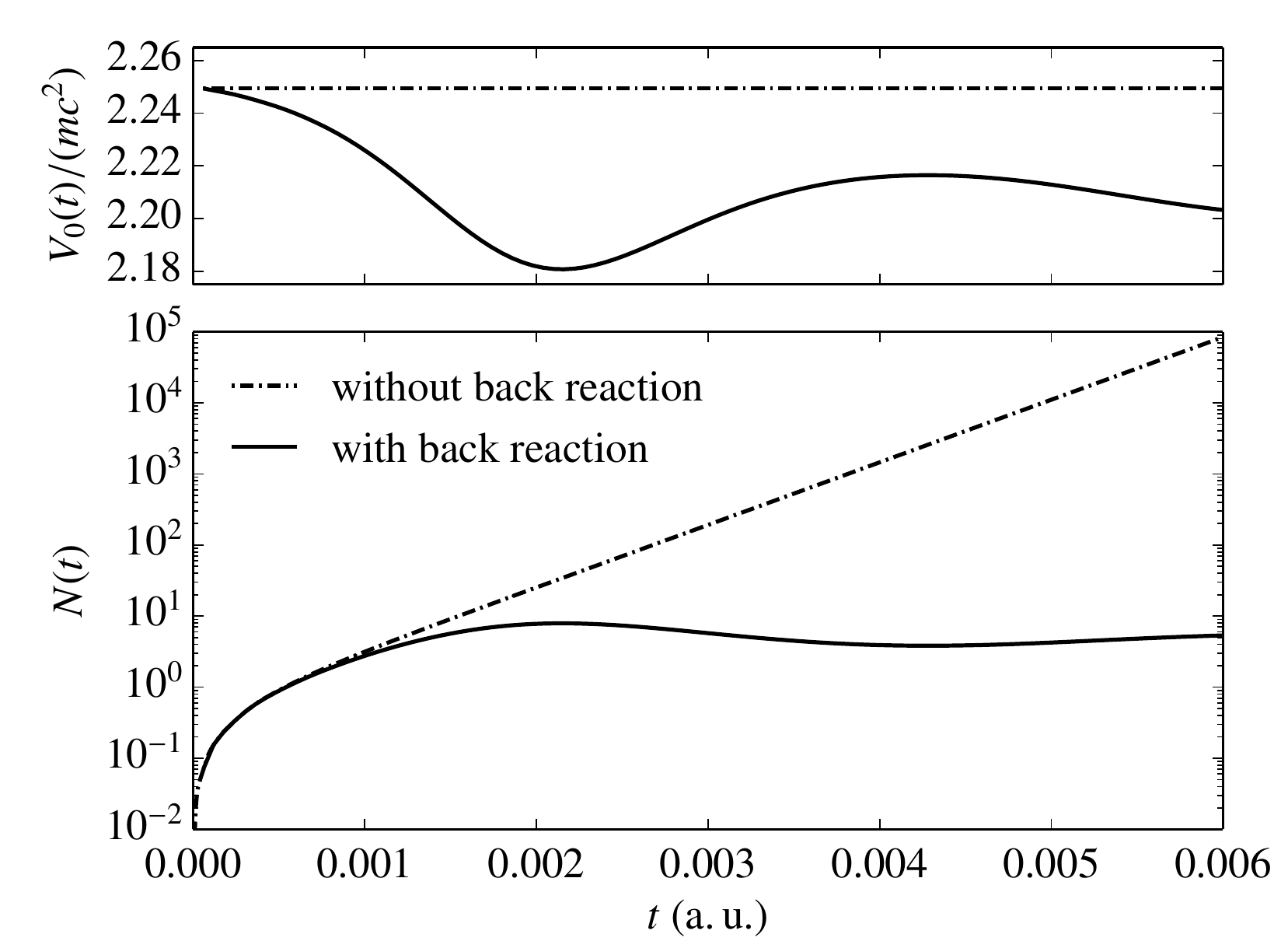}
  \caption{The created particle number $N(t)$ and the depth of the
    potential $V_0(t)$ with and without back reaction as functions of
    time $t$. The parameters used here are the same as in regime~IV of
    Fig.~\ref{fig:created_particles}}
  \label{fig:br_creation}
\end{figure}

Incorporating this change of the energy of the external field affects
significantly the pair-creation process as shown in
Fig.~\ref{fig:br_creation}.  Here we have fixed the spatial distribution
of the external field and the change of the energy is only reflected by
a time-dependent potential strength $V_0(t)$, which is dynamically
adjusted such that the energy of the scalar potential equals
Eq.~\eqref{eq:field_energy}.  There is no sustained pair creation if
back reaction is included in our model.  Initially, the number of
particles grows but it turns soon into an oscillatory behavior.  Due to
the created particles and back reaction, the potential can become
subcritical.  In this case, the anti-bosons do not have enough kinetic
energy to escape from the interaction region and will therefore
annihilate with bosons and, consequently, the external field will become
supercritical again and create another pair.  This process lasts forever
as the external field oscillates between supercritical and subcritical.

As an alternative to our purely phenomenological approach, the inclusion
of the back reaction could be implemented on a more fundamental level,
which would extend the theoretical description significantly.  In the
case of the related fermionic pair creation, some progress in this
direction has been reported recently
\cite{Hebenstreit:Berges:etal:2013:Real-Time_Dynamics,Steinacher:Wagner:etal:2014:Impact_of,Kasper:etal:2015:Schwinger_pair_production},
where the back reaction of the created electron-positron pairs on the
electromagnetic field has been taken into account by coupling the Dirac
equation with the Maxwell equation under the assumption that the force
field is classical.  However, such an approach lies beyond the scope of
this paper and, furthermore, it is presently not clear how this
procedure could be applied to bosonic systems.

\section{Conclusions}
\label{sec:Conclusions}

In this paper, we studied the role of pseudodegenerate bound states on
pair creation in bosonic systems.  The coalescing of particle and
antiparticle bound states when increasing the potential strength is
known as the Schiff-Snyder-Weinberg effect.  These coalesced states have
complex energy eigenvalues due to the pseudo-Hermiticity of the
Klein-Gordon Hamiltonian.  By employing the biorthogonal left and right
eigenvectors of the Hamiltonian, these pseudodegenerate states can be
used as basis sets to quantize the bosonic matter field.  The problem
that states with complex energy have zero norm does not occur when one
discriminates thoroughly between left and right eigenvectors.  When
distinct bound particle and antiparticle states merge into pseudo
degenerate states, they trigger pair creation from the vacuum.  The mean
particle number obeys an exponential growth law in time.  The
characteristic parameter of this exponential behavior equals to twice
the absolute value of the imaginary part of the energy of the pseudo
degenerate states.

In addition to the simple scalar potential well, we also studied a field
configuration that consists of a superposition of an electric and a
magnetic field.  Our findings are contrary to the common belief that the
Schiff-Snyder-Weinberg effect can only occur for short-range potentials.
The scalar and the vector potential used here are both long range,
nevertheless the coalescing of bound states is found here, too.  The
magnetic field given by the vector potential is essential for this
generalized Schiff-Snyder-Weinberg effect.  The generalized
Schiff-Snyder-Weinberg effect induces pair creation from the vacuum,
which leads to an exponential increase of the particle number in time.
The exponential creation rate is related to the imaginary part of the
pseudodegenerate energies as for the standard Schiff-Snyder-Weinberg
effect.

Exceeding some critical potential strength, not only are the bound
states degenerate with respect to another bound state or states in the
continuum but also the lower and upper continuum bands become degenerate
to each other as the potentials are long range.  The resulting
continuum-continuum overlap induces a different mechanism for pair
creation that (by itself) would lead to a permanent linear growth of the
particle yield.  As a result of both mechanisms, this
continuum-continuum transition plays a minor role for the overall
long-time behavior, which shows exponential growth.

The characteristic of the generalized Schiff-Synder-Weinberg effect is
that pseudodegenerate states with complex energy emerge and dissolve
again as a function of some potential parameter.  Consequently, a
remarkable feature of the generalized Schiff-Synder-Weinberg effect is
that the growth rate of the number of created particles is not a
monotonous function of the potential strength.  This means, a strong
field may create fewer particles than some weaker field.

In contrast, a Schiff-Snyder-Weinberg-like effect has never been found
in the fermionic systems and the Dirac Hamiltonian usually features the
avoided crossing mechanism, when two states approach each other.  The
comparison of the pseudodegeneracy and the avoided crossing may give us
some insights about the physical meaning of the spatial density of the
states.  More systematic studies of the pseudodegeneracy in the
framework of quantum field theory are needed.

Pair creation via strong static fields has not been experimentally
verified for either fermions or bosons.  While we expect that due to
their smaller rest masses electron-positron pairs might be observed in
the near future, the lowest massive charged bosons (such as
$\pi$~mesons) are heavier and require therefore even stronger external
fields for their production.  However, the theoretical study of bosonic
systems in extreme parameter regimes allows us to explore the limits of
quantum theories.  To stress this we would like to finish this paper by
drawing the reader's attention to the following fundamental issue.  The
Hamiltonian of the Dirac equation for an electron in a Coulomb potential
loses its mathematical property of being self-adjoint
\cite{Ruf:Muller:etal:2011:Numerical_signatures} if the nuclear charge
exceeds~137.  Therefore, it is often conjectured that in this particular
regime this theoretical framework has reached the principal limits of
its applicability and a collapse of the vacuum is postulated
\cite[Chap.~9]{Greiner:1997:Relativistic_Quantum_Mechanics}.  As we have
discussed in the present work there is also a parameter regime where a
discrete subset of the spectrum of the Klein-Gordon Hamiltonian becomes
complex.  We should point out that it is possible that the Klein-Gordon
theory becomes physically meaningless in this regime, similar to the
conjectures regarding the Dirac equation, or requires a modified
interpretation.  However, a definite answer to this extremely
fundamental question cannot be obtained within any theoretical framework
alone and would clearly require experimental data.

\acknowledgments

This work has been supported by the NSF and the National Natural Science
Foundation of China under Grant No.~11529402.  R.~G.\ acknowledges the
warm hospitality during his visit at the Max-Planck-Institut f{\"u}r
Kernphysik in Heidelberg.

\bibliography{SSW_effect_pair_creation}

\end{document}